\documentclass{aa}  
\usepackage{graphicx}
\usepackage{txfonts}
\usepackage{soul}
\usepackage{amsmath,tabu}
\usepackage{relsize}
\usepackage{multicol}
\usepackage{multirow}
\usepackage{booktabs}
\usepackage{dirtytalk}
\usepackage{subcaption}
\usepackage{threeparttable}
\usepackage{hyperref}
\hypersetup{colorlinks,linkcolor=blue,citecolor=blue,urlcolor=blue}
\usepackage{array}
\usepackage{natbib,twoopt}
\bibpunct{(}{)}{;}{a}{}{,}             
\defcitealias{Gosnell_2015}{G15}
\defcitealias{Ferraro06}{F06}

\begin{document}

\authorrunning{E. Reggiani}
\titlerunning{Detection of a  white dwarf orbiting a Carbon-Oxygen depleted blue straggler in 47 Tuc}

\title{{Detection of a  white dwarf orbiting a Carbon-Oxygen depleted blue straggler in 47 Tucanae}\thanks{Based on observations with the
    NASA/ESA HST, obtained under programme GO 15914 (PI: Lanzoni). The
    Space Telescope Science Institute is operated by AURA, Inc., under
    NASA contract NAS5-26555.}}

\author{Elisabetta Reggiani\inst{1,2,3}\thanks{\email{elisabetta.reggiani@unifi.it}},
        Mario Cadelano\inst{3,4},
        Barbara Lanzoni\inst{3,4},
        Francesco R. Ferraro\inst{3,4},
        Maurizio Salaris\inst{5,6},
        Alessio Mucciarelli\inst{3,4}
        }

\institute{Dipartimento di Fisica e Astronomia, Università di Firenze, Via G. Sansone 1, 50019 Sesto Fiorentino, FI, Italy\
    \and
        INAF - Osservatorio Astrofisico di Arcetri, Largo E. Fermi 5, 50125 Firenze, Italy
    \and
        Dipartimento di Fisica e Astronomia, Università degli Studi di Bologna, Via Gobetti 93/2, I-40129 Bologna, Italy   
    \and
        INAF, Osservatorio di Astrofisica e Scienza dello Spazio di Bologna, Via Gobetti 93/3, I-40129 Bologna, Italy
    \and
        Astrophysics Research Institute, Liverpool John Moores University, Liverpool L3 5RF, UK 
    \and    
        INAF - Osservatorio Astronomico D'Abruzzo, Via Mentore Maggini s.n.c., 64100 Teramo, Italy        \\
    }

\abstract
{We have employed deep far-UV observations secured with the Solar Blind Channel of the Advanced Camera for Surveys onboard the Hubble Space Telescope to search for hot companions to five blue stragglers stars (BSSs) showing significant surface depletion of carbon (C) and oxygen (O), in the Galactic globular cluster 47 Tucanae. Such a chemical pattern has been interpreted as the chemical signature of the mass transfer formation process for the observed blue stragglers. The mass transfer origin is also expected to leave a ``photometric signature" in the form of a UV-excess, as the stripped core of the donor star should be observable as a white dwarf (WD) companion orbiting the newborn BSS. We found strong evidence for the presence of a hot ($T>20000$ K) WD companion to one of the investigated BSS, indicating that it likely formed through mass transfer less than $\sim 12$ Myr ago. This is the first simultaneous evidence of the chemical and the photometric signatures of the mass-transfer formation channel. The lack of evidence for a hot companion to the other investigated blue stragglers is consistent with the expectation that the photometric signature (as well as the chemical one) is a transient phenomenon.
}
\maketitle
\section{Introduction} 
\label{intro}
Blue straggler stars (BSSs) are core hydrogen-burning objects more massive than main sequence (MS) and post-MS stars. They are observed in all stellar environments, ranging from globular clusters \citep[GCs; e.g.][]{Sandage_1953, Ferraro_2003,Ferraro_2018,  piotto+04, leigh+07, leigh+13, knigge+09}, to open clusters \citep[e.g.][]{mathieu+09, rain+21, Rain_2024}, the Galactic field \citep[e.g.][]{Preston_2000, clarkson+11}, and dwarf spheroidal galaxies \citep[e.g.][]{momany+07, Mapelli_2009}.  The observational and theoretical evidence accumulated so far demonstrates that BSSs are more massive than the other stars populating a cluster (see \citealt{Shara_1997,gilliland+98,Fiorentino2014,Raso2019})
and for this reason they are invaluable probes of star cluster internal dynamical evolution \citep{Ferraro_2012,Ferraro_2018,ferraro+19, ferraro+23a, alessandrini2016, Lanzoni2016, khushboo+23}, possibly also offering the possibility to pinpoint and date the occurrence of core collapse (\citealt{Ferraro_09, dalessandro+13, portegies+19, beccari+19, ferraro+20, cadelano+22}; see also \citealp{simunovic+14} and \citealp{raso+20}).

\begin{figure*}
     \centering
     \includegraphics[scale = 0.15]{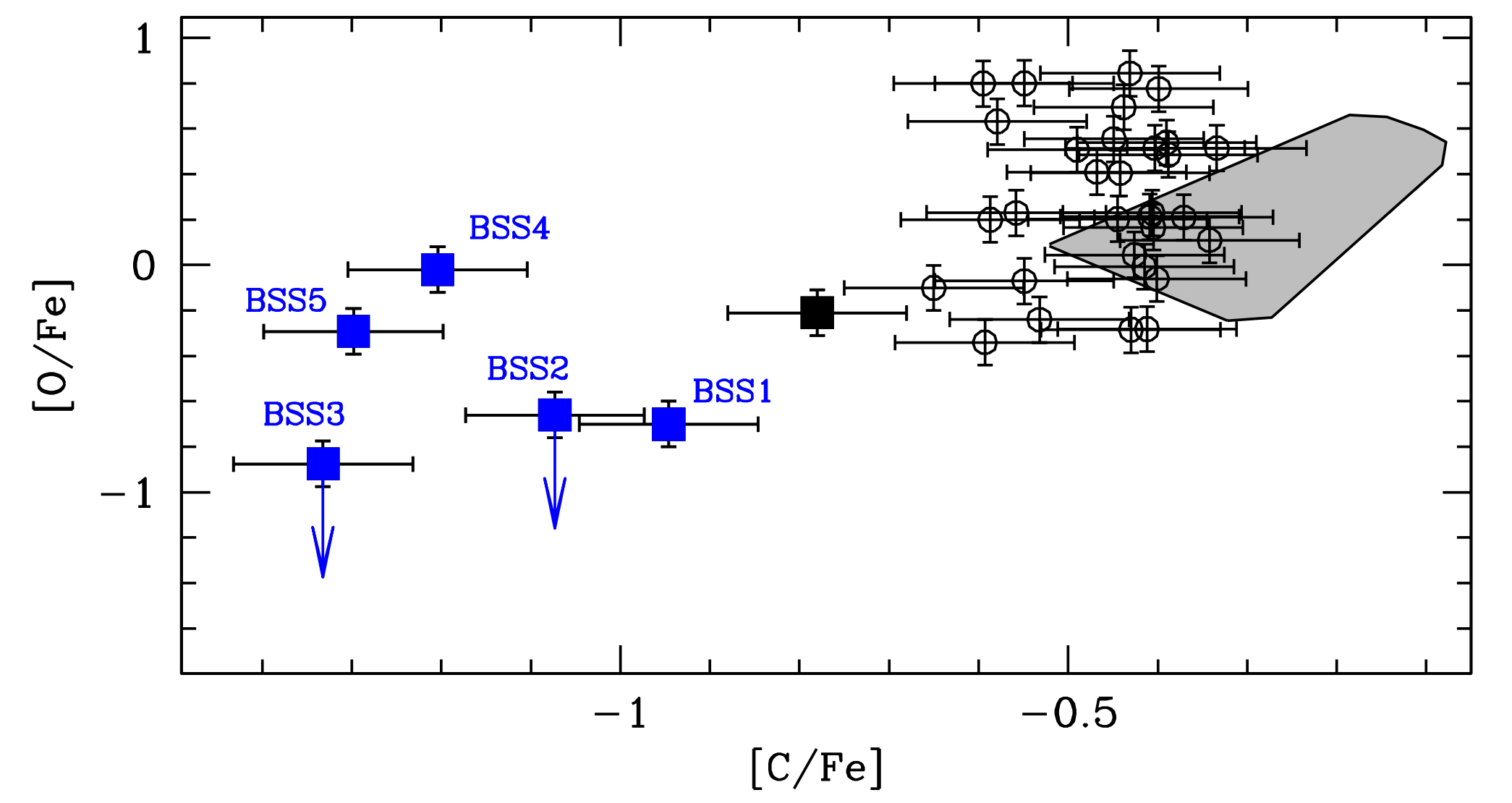}
     \caption{Position in the [O/Fe]$-$[C/Fe] diagram of the 43 BSSs (squares and circles) and MS-TO stars (gray shaded region) of 47 Tucanae discussed in \citet{Ferraro06}. The blue squares mark the 5 targets of this study. 
     }
\label{fig:COdep}
\end{figure*}

Despite the astrophysical importance of these objects, their formation mechanisms are not fully understood yet. To account for their mass being larger than the MS turn-off (TO) value in stellar systems that experienced just a single star-formation event, formation channels involving mass-enhancement processes have been proposed: direct stellar collisions \citep{HD_1976, Sills2005}, stellar mergers occurring on secular timescales through angular momentum loss or Kozai oscillations \citep[see, e.g.,][]{andronov+06, perets_fabrycky09}, and mass transfer (MT) activity in binary systems \citep{McCrea_1964}. While stellar collisions are expected to be more frequent in highly crowded environments (e.g. \citealt{Davies_2004}), binary and triple mergers and the MT formation process should be dominant in low-density environments (e.g. \citealt{Sollima_2008, knigge+09, mathieu+09}). The relative efficiency of the various mechanisms is however still unclear, also because disentangling the different kinds of BSSs based on their observed properties is very challenging. However, \citet{ferraro+25} recently found evidence that the vast majority of BSSs detected in Galactic GCs have a binary-related origin (see also \citealt{knigge+09}), thus identifying MT as the most effective formation channel. In this scenario a MS star is bound in a binary system and accretes matter from a companion that has filled its Roche lobe. At the end of the MT process the accreting MS star becomes a BSS, while the core of the stripped companion should be observable as a helium or a carbon-oxygen white dwarf (WD), depending on the evolutionary stage of the donor (sub-giant or red giant branch in the former case, asymptotic giant branch in the latter).  The matter settled on the BSS surface should come from the inner regions of the donor star, where it was processed by thermonuclear reactions, hence chemical anomalies should be detectable in a BSS atmosphere \citep{Sarna1996}. In old globular clusters, where stellar masses and metallicities are such that s-process enhancements are not expected to occur (but see, e.g., \citealp{jorissen+19} and \citealp{escorza+19} for the case of Barium dwarfs), lower abundances of carbon (C) and oxygen (O) are expected on the surface of MT-BSSs,
compared to normal cluster stars, since these elements are depleted during the CNO cycle in hydrogen-shell burning. Conversely, normal chemical patterns are predicted for collisional BSSs \citep{Lombardi1995}. Thus, in principle, two main observable signatures should characterise the outcome of the MT process: 
\begin{enumerate}\itemsep2pt \parskip0pt \parsep0pt
    \item MT-BSSs should have WD companions (photometric signature)
    \item MT-BSSs should show C and O depletion on their surface (chemical signature)
\end{enumerate}
The photometric signature has been first identified in the open cluster NGC~188, where significant UV excess has been detected at the position of 7 BSSs, and interpreted as evidence for the presence of hot WD companions, as expected in the case of the MT formation channel (\citealt{Gosnell_2015} -- hereafter \citetalias{Gosnell_2015}; see also \citealp{Gosnell_2014}). Similar results have been subsequently found in other open and globular clusters \citep[e.g.][and references therein]{subramaniam+16, sahu+19, dattatrey+23}.


As for the chemical signature, measuring the chemical composition of BSS is not an easy task due to the effect of radiative levitation that alters the surface chemical abundances
in BSSs hotter than $\sim8000$ K \citep{lovisi+12}.  
Nevertheless, the chemical analysis of BSSs has been performed with some promising results in a couple of GCs. In particular,  high-resolution spectroscopic observations in the GC 47 Tucanae   (\citealt{Ferraro06}-- hereafter \citetalias{Ferraro06}) led to the identification of 5 BSSs with significant CO-depletion with respect to the dominant population (see Figure \ref{fig:COdep}; \citetalias{Ferraro06}). Moreover, some O depletion was also detected in 4 BSSs (out the five ones not affected by radiative levitation) located along the red sequence in M30 (see \citealt{Lovisi2013}).

To search for the presence of both photometric and spectroscopic signatures of the MT formation process in the same object, here we present the results of a photometric investigation in the UV domain aimed at assessing the possible presence of a WD companion to the 5 BSSs with C-O depletion detected in 47 Tucanae by \citetalias{Ferraro06}.
%
In Section \ref{sec:dataanalysis} we present the observations and the adopted data reduction procedure. Section \ref{sec:isochrones} provides first estimates of the physical parameters of the target BSSs as obtained from the comparison with isochrones in the CMD. In Section \ref{sec:sed} we describe the procedure adopted to build the observed spectral energy distributions (SEDs) of the five BSSs and compare them with theoretical SEDs computed for isolated stars. Section \ref{sec:wd_fitting} demonstrates the necessity of a hot WD companion to BSS4, and Section \ref{sec:conclusions} presents the discussion and conclusions of the obtained results.

\begin{figure*}
     \includegraphics[width=0.98\textwidth]{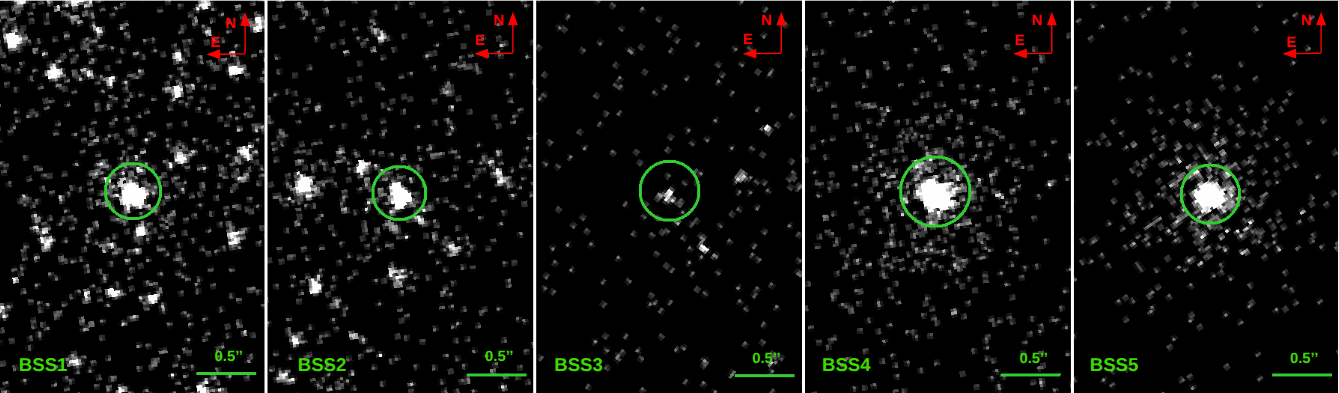}
     \caption{Drizzled ACS/SBC images in the F165LP filter of the five BSSs investigated in this study. The positions of the 5 targets are marked by green coloured circles: from left to right, BSS1, BSS2, BSS3, BSS4, and BSS5. The scale is indicated in the bottom left corner of each frame.}
     \label{sbc_images}
\end{figure*}

\section{Observations and data reduction}
\label{sec:dataanalysis}
To search for possible WD companions to the five CO-depleted BSSs in 47 Tucanae, we 
took advantage of the superb far-UV capabilities of the Solar Blind Channel (SBC) of the Advanced Camera for Surveys (ACS).
The photometric dataset consists of a set of ACS/SBC images acquired under GO15914 (P.I.: B. Lanzoni) 
through filters F140LP, F150LP, F165LP. 
The two BSSs located in the innermost portion of the cluster (namely, BSS1 and BSS2) have been observed in the same pointing, thus a total of 4 pointings were necessary to sample the 5 targets. The dataset consists  of 16 images for each of the 4 pointings. Specifically, 6 exposures ($\sim$740 s each) were secured through the F140LP filter;   
7 exposures ($\sim$740 s each) in filter F150LP
and 3 exposures ($\sim$750 s each) in filter F165LP.
The five target as they appear in the SBC images obtained in the F165LP filter are shown in Fig.\ref{sbc_images}.
\begin{table*}
\caption{ACS/SBC far-UV magnitudes of the 5 targets}
\label{tab2}
\centering
\setlength{\tabcolsep}{7.pt}
\renewcommand{\arraystretch}{1.6}
    \normalsize
    \begin{tabular}{c c c c c c}
    \hline\hline
    ID & $\alpha$ (J2000) & $\delta$ (J2000) &  $ m_{F140LP}$   &  $m_{F150LP}$    &    $m_{F165LP} $      \\
    \hline
    BSS1 & 00 24 07.89 & -72 05 01.77 &   $21.72 \pm 0.08$ &   $21.23\pm 0.05$ &    $20.25 \pm 0.09$ \\   
    BSS2 & 00 24 11.69 & -72 04 44.11 &   $22.85 \pm 0.12$ &   $22.34\pm 0.11$ &    $20.20 \pm 0.15$ \\  
    BSS3 & 00 24 04.39 & -72 00 35.90 &   ... &   ... &    $23.39 \pm0.23$  \\  
    BSS4 & 00 23 14.63 & -72 07 37.06 &   $19.11 \pm 0.04$ &   $19.23\pm 0.03$ &    $19.07 \pm 0.03$ \\  
    BSS5 & 00 26 10.47 & -72 11 07.64 &   $20.85 \pm 0.06$ &   $20.63\pm 0.07$ &    $19.19 \pm 0.01$ \\  
    \hline
    \end{tabular}
    \label{uvmag}
\end{table*}
 The raw data were retrieved from the Archive and corrected for flat field and bad pixels. Geometrical distortions were corrected by applying the Pixel Area Map that also allows the removal of inhomogeneities in the relative sensitivity of the MAMA detector pixels.

The photometric analysis for the BSSs in the external regions has been performed by using the IRAF package DAOPHOT II \citep{Stetson1987} via aperture photometry, with a procedure similar to that used in \citet{Nine2023}. We extracted count rates using an aperture radius of 40 pixels (corresponding to 1") and then applied an encircled energy fraction correction to our count rates following the results of \citet{Avila_Chiaberge_2016}, who found the encircled energy correction factor to be approximately 0.92 at a radius of 1" in each of F140LP, F150LP, and F165LP. For BSS3, the emission at the star's coordinates was so low that we were able to obtain a value only for the redder filter, F165LP.
The presence of nearby sources in the pointing of BSS1 and BSS2 required us to use a different approach for the photometric analysis of their images, recurring to a PSF fitting procedure performed through DAOPHOT II following the prescriptions in \citet{cadelano20b,cadelano20,chen_nature}.
The instrumental magnitudes were then calibrated in the VEGAMAG system using the zero points obtained with the ACS zero point calculator \footnote{\href{https://acszeropoints.stsci.edu/}{https://acszeropoints.stsci.edu/}}.
The ACS/SBC is known to have a red leak above 2000 \AA\ that significantly affects solar or later-type stars\footnote{\href{https://hst-docs.stsci.edu/acsihb/chapter-5-imaging/5-5-ultraviolet-imaging-with-the-sbc}{ACS Instrument Handbook}}. The magnitude corrections obtained for our targets following the results reported in \citet{Nine2023} are included in the values listed in Table \ref{uvmag}, which lists the derived magnitudes of the five targets in each filter.

\begin{figure*}
     \centering
     \includegraphics[scale = 0.16]{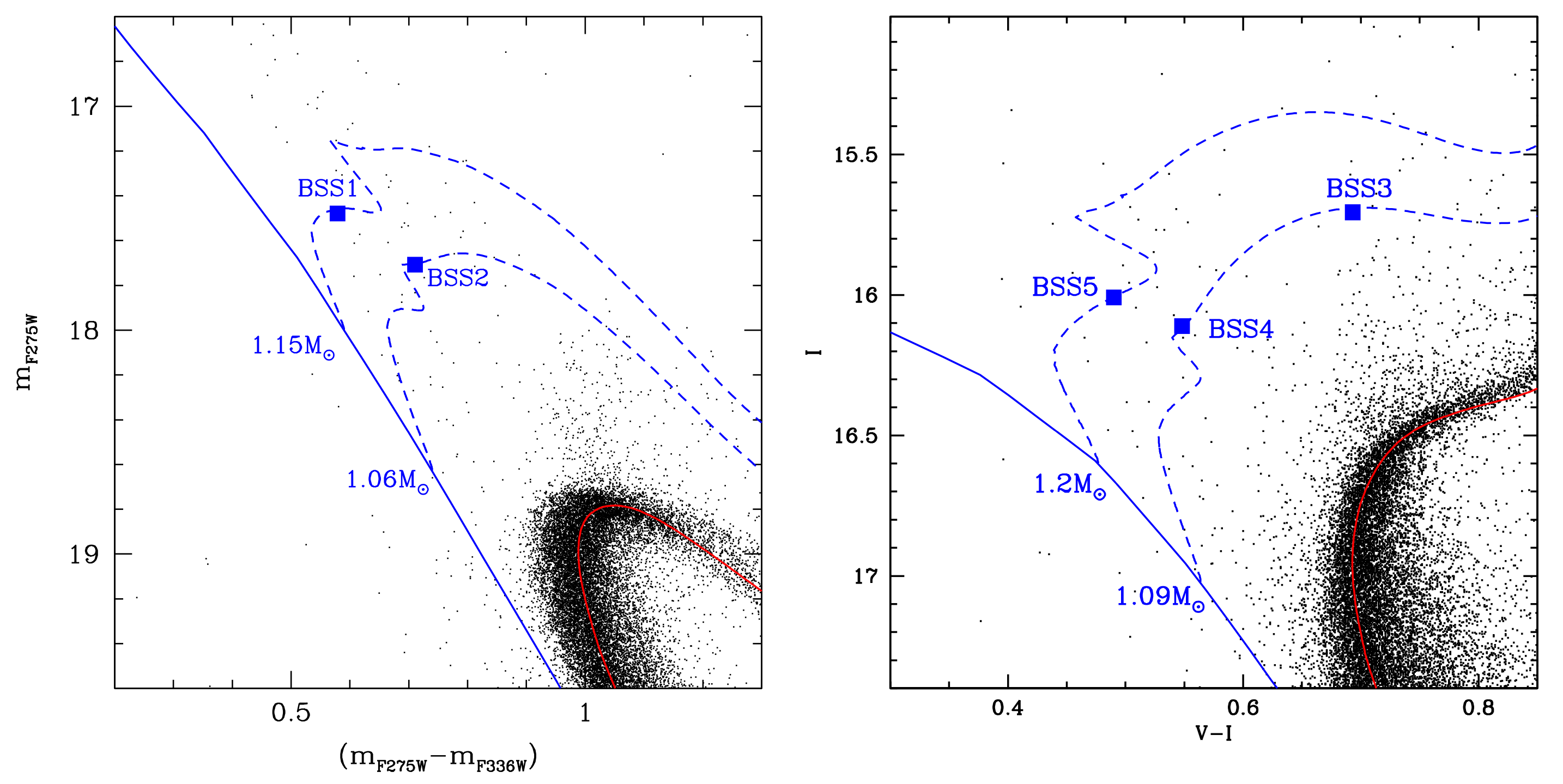}
     \caption{Near-UV CMD (left panel) and optical CMD (right panel) of 47 Tucanae with the 5 targets of this study highlighted as blue squares. The 12 Gyr old BaSTI \citep{Pietrinferni_2021} isochrone well reproducing the MS-TO region is shown in both CMDs as a red line. The corresponding 40 Myr old isochrone, assumed to be representative of the zero-age MS location, is plotted as a solid blue line. The dashed blue lines are the evolutionary tracks that best reproduce the observed positions of the targets in these CMDs.}
\label{fig:CMD}
\end{figure*}

\section{First-guess estimate of the target physical parameters}
\label{sec:isochrones}
First-guess estimates of the physical properties (such as temperature and mass) of the investigated targets can be derived from the comparison between their position in the CMD and theoretical models of individual stars. To this end, we considered the $(m_{\rm F275W}, m_{\rm F275W}-m_{\rm F336W})$ UV CMD (see the left panel in Figure \ref{fig:CMD}) for the two innermost targets (BSS1 and BSS2), which lie in the field of view of the HST UV survey \citep{Nardiello2018}, while we used the optical ($I, V-I$) photometry from \citet{Stetson2019} for the remaining BSSs (see the right panel in Figure \ref{fig:CMD}). The CMD position of BSS3 
suggests that this could be an evolved BSS. As theoretical models, we used a set of isochrones and evolutionary tracks from the BaSTI database \citep{Pietrinferni_2021} computed for [Fe/H]$=-0.7$ and an $\alpha$-enhanced [$\alpha$/Fe]=0.4 mixture, in agreement with the chemical abundances measured in 47 Tucanae \citep[e.g.][]{carretta09,mcwilliam+08}. 
We adopted the distance modulus and reddening quoted in \citet[][2010 version]{harris96}, and applied small shifts in magnitude and colour to optimise the match to the data. As shown in Fig. \ref{fig:CMD}, a 12 Gyr old isochrone nicely reproduces the MS-TO region of the cluster in both CMDs. We used a very young (40 Myr old) isochrone to locate the theoretical zero-age MS, and we extracted a set of BaSTI evolutionary tracks for stellar masses ranging from 1.0 to 1.5 $M_\odot$, stepped by 0.01 $M_\odot$, selecting those that best reproduce the position of each target in the adopted CMDs. We found that the targets are consistent with masses ranging from 1.06 (BSS2) up to 1.2 $M_\odot$ (BSS5). Interestingly, BSS3 and BSS4 are located along the same evolutionary track, corresponding to a stellar mass of 1.09 $M_\odot$.   
The nice matches allowed us to extract the physical parameters of each target directly from the best-fitting evolutionary track, which are  
listed in Table~\ref{photo_estimates}. We remind that caution is needed when estimating BSS masses from the comparison with evolutionary tracks computed for normal, single stars. However, as shown in \citet{Raso2019}, a reasonable agreement is found between these estimates and those obtained from the fit to the observed SED, thus suggesting that the former can be used at least for first-guess values.

\begin{table}[h!]
\caption{First-guess physical parameters of the 5 targets as derived from the comparison with stellar tracks.}
\label{tab2}
\centering
\setlength{\tabcolsep}{5.pt}
\renewcommand{\arraystretch}{1.5}
    \normalsize
    \begin{tabular}{c c c c c c}
    \hline\hline
    ID &  Mass/$M_{\odot}$ &  Temperature (K)  &  $L/L_{\odot}$  & $R/R_{\odot}$ & $\log g$   \\
    \hline
    BSS1 &  1.15 & 6760 & 3.6 & 1.45 & 4.18 \\   
    BSS2 &  1.06 & 6490 & 3.4 & 1.49 & 4.12 \\  
    BSS3 &  1.09 & 5960 & 5.2 & 2.18 & 3.80 \\ 
    BSS4 &  1.09 & 6550 & 3.9 & 1.56 & 4.10\\  
    BSS5 &  1.20 & 6810 & 4.5 & 1.55 & 4.14 \\  
    \hline
    \end{tabular}
    \label{photo_estimates}
\end{table}

\section{Constructing the Spectral Energy Distribution}
\label{sec:sed}
The presence of a hot WD companion orbiting a BSS is expected to produce a far-UV flux significantly exceeding the one expected for an isolated BSS of given effective temperature. As a first step, we have built the observed SED of each BSS by combining the far-UV magnitudes determined in this work, with previous measurements obtained at different wavelengths. The photometry available for each target is reported in Table \ref{mag_table} and includes near-UV and optical HST data \citep[from][]{Brown_2009, Cadelano2015, Nardiello2018, Pantoja_2018, Rivera_Sandoval_2020} for the two innermost BSSs, GAlactic Evolution EXplorer \citep[GALEX;][]{Dalessandro2012}, UBVRI ground-based photometry \citep{Stetson2019}, GAIA \citep{gaia_23}, and Two Micron All Sky Survey \citep[2MASS;][]{Skrutskie+06} data for some of the most external targets. 

Each 
magnitude has been de-reddened assuming $E(B-V) = 0.04$ \citep[][2010 version]{harris96} and the extinction curve by \citet{Cardelli1989}. Finally, we converted magnitudes into observed fluxes by using the following relation:
\begin{equation}
    F = \text{\textit{photflam}}\cdot10^{-\mathlarger{\frac{m_{\lambda,corr}}{2.5}}}
\end{equation}
where $m_{\lambda,corr}$ is the de-reddened magnitude and $photflam$ is the inverse sensitivity (i.e.,  the flux of a source with constant $F_{\lambda}$, which produces a count rate of 1 electron per second).
\begin{figure*}
    \centering
    \includegraphics[width=\textwidth]{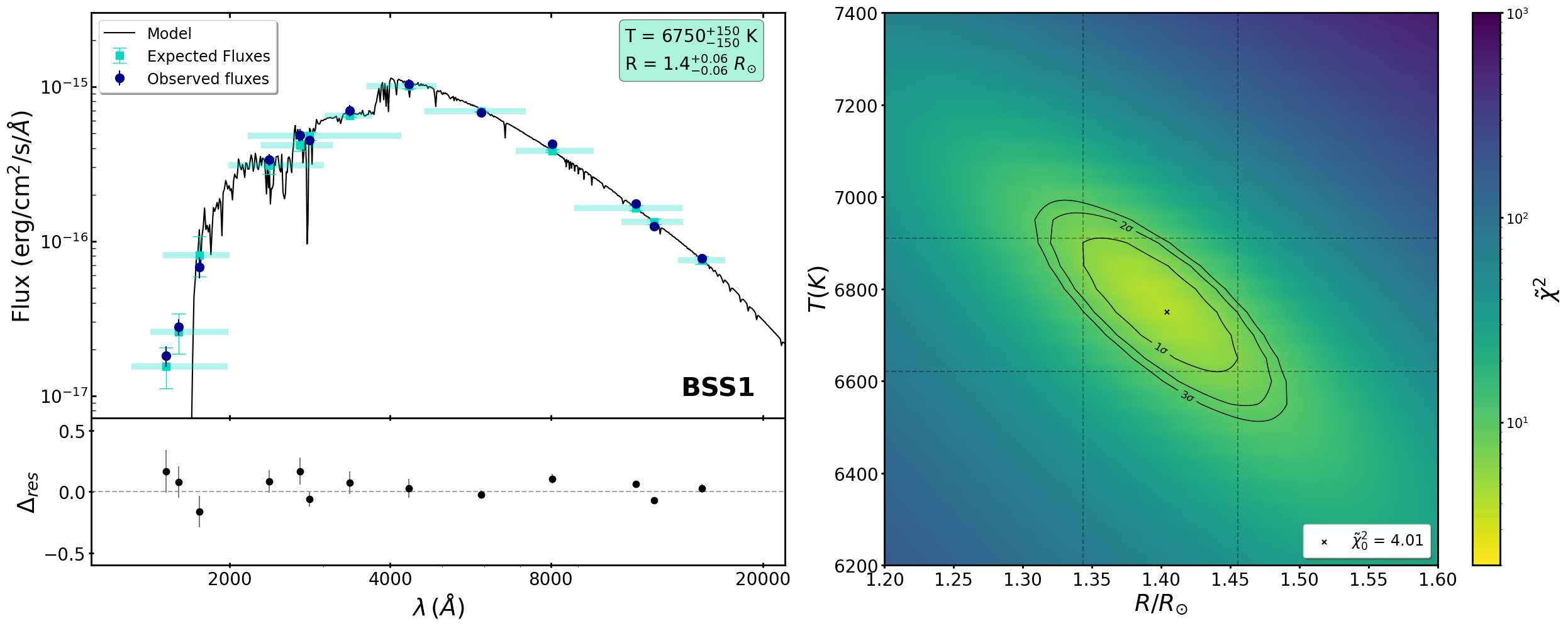}
    \caption{\textit{Left Panel:} Observed SED of BSS1 (blue circles) compared to the expected fluxes (cyan squares) computed from the convolution of the best-fit synthetic SED (black line) with the adopted photometric filters. The uncertainties on the observed fluxes are marked with vertical errorbars (unless they are smaller than the size of the blue circles). The horizontal cyan-shaded rectangles mark the wavelength width of each photometric filter. For the sake of clarity, they have been associated to the expected fluxes only but, of course, they also hold for the observed points. The vertical errorbar of the cyan squares corresponds to the 1$\sigma$ error on the expected fluxes. The best-fit surface temperature and radius are labelled together with their $1\sigma$ uncertainty in the top-right corner legend. The residuals between the observed and the expected fluxes are plotted in the lower panel. \textit{Right Panel:} $\chi^2$ map for BSS1 showing the distribution of $\chi^2$ values (colour-coded as in the side bar) obtained for all the explored combinations of R and T. The black lines refer to the $1\sigma$, $2\sigma$ and $3\sigma$ $\chi^2$ contours from the minimum values ($\chi_0^2$), which is marked with a black cross and labelled in the bottom-right corner.}
     \label{bss1_fit}
 \end{figure*}
The values of the extinction law and the inverse sensitivity we used for each filter are reported in Table~\ref{photflams}.

\begin{figure*}
    \centering
    \includegraphics[width=\textwidth]{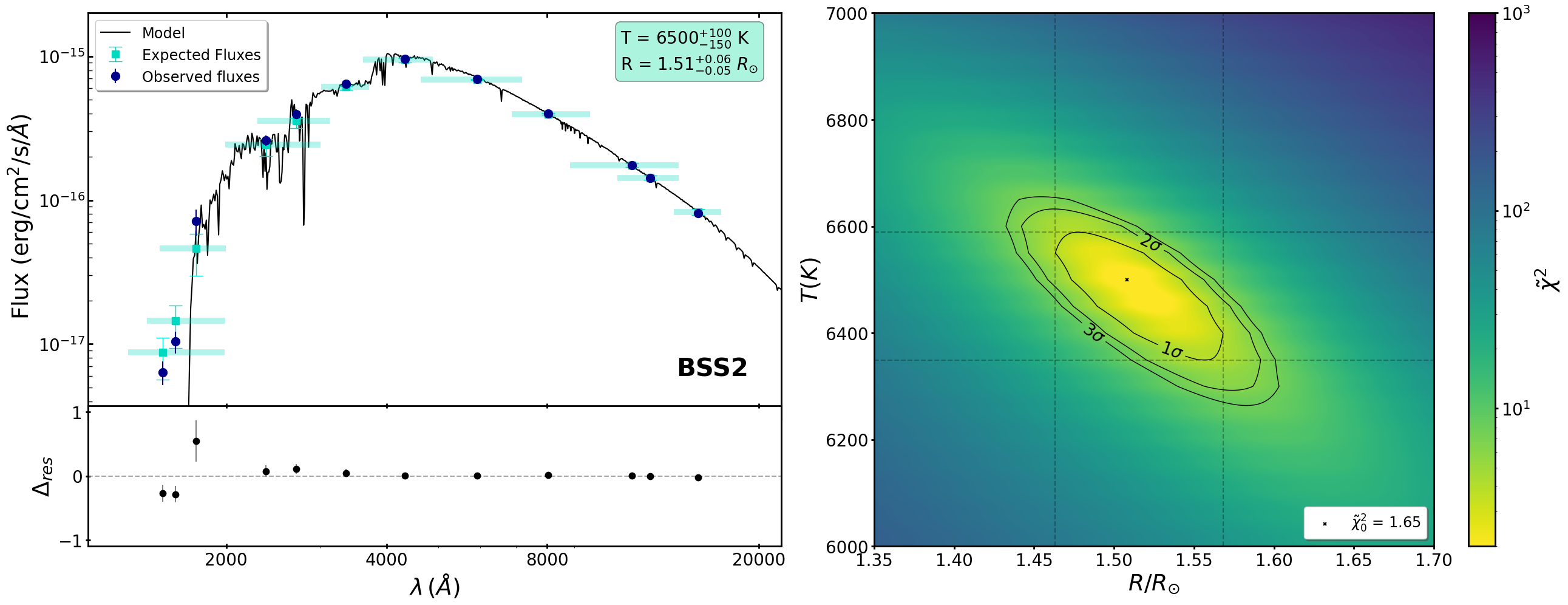}
     \caption{Same as in Figure~\ref{bss1_fit}, but for BSS2.}
     \label{bss2_fit}
 \end{figure*}

The theoretical SEDs have been determined using a new grid of stellar fluxes (A. Mucciarelli et al. in prep.) calculated using the ATLAS9 code \citep{Kurucz+05}.
All the theoretical fluxes are calculated adopting [M/H]=-0.75 and an alpha-enhanced chemical mixture ([$\alpha$/Fe]=+0.4 dex), with effective temperatures ranging from 5000 to 9000 K in steps of 50 K, and surface gravity $3.5 \leq \log{g} \leq 4.7$ with steps of 0.1 dex. To compare the observed flux in every filter with the theoretical one ($f_\lambda$), we determined
the theoretical apparent flux ($F_\lambda$), by applying the following relation:
\begin{equation}
    F_{\lambda} = \Bigg(\frac{R}{D}\Bigg)^2 f_{\lambda}
\end{equation}
where $R$ is the stellar radius and $D$ is the distance of the source \citep[approximated as the distance of 47 Tucanae; we adopted $4.55$ kpc;][]{Ferraro1999}.
The theoretical flux in each filter has been calculated by convolving the 
synthetic spectra with the appropriate bandpasses using the python module \texttt{pysynphot}\footnote{Lim, P. L., Diaz, R. I., \& Laidler, V. 2015, PySynphot User’s Guide \citep[Baltimore, MD:][]{pysynphot}}. Since the SED is insensitive to surface gravity variations in the sampled wavelength range, for the fitting procedure we fixed $\log{g}$ to the values estimated from the photometry (Table \ref{photo_estimates}), leaving temperature and radius to vary between 5000 and 9000 K, and between $0.5 R_\odot$ and $2.5 R_\odot$, respectively.
The best-fit model was found by means of a $\chi^2$ minimisation procedure, which identified the temperature-radius combination needed to best match the observed SED. The comparison between the observed and the best-fit synthetic SEDs is shown in the left panel of Figs.~\ref{bss1_fit}-\ref{bss5_fit}.
For each star, we built a $\chi^2$ map (see the right panels in the same figures) and used the limits of the contours at 1$\sigma$ to estimate the errors associated to the best-fit temperature and radius. The results of the SED-fitting procedure are discussed separately in the next sections for each investigated BSS.

\subsection{BSS1}
The best-fit SED model of BSS1 reproduces all the available photometric data with a surface temperature that agrees well with the photometric estimate. The 1$\sigma$ contour in the $\chi^2$ map (right panel) covers a small range of values for the stellar radius, whose best fit is consistent within the errors with the value listed in Table \ref{photo_estimates}. 
Since the theoretical SED reproduces all the observed photometric points with reasonable values of the stellar parameters, the presence of a hot companion can be safely ruled out. 

\subsection{BSS2}
Similarly to BSS1, also for BSS2 the best-fit SED model well reproduces both visible and  UV observations, with a surface temperature consistent with that reported in Table~\ref{photo_estimates}, and with a small spread in the radius values. For this reason, the presence of a hot companion can be safely ruled out also for this star.
 
\subsection{BSS3}
As discussed in Sect. \ref{sec:isochrones}, this BSS lies in a rather red position in the CMD (right panel of Fig. \ref{fig:CMD}), and since it emits no significant flux in the SBC filters (Fig. \ref{sbc_images}), we used only the F165LP data point in the fitting procedure.
\begin{figure*}
    \centering
    \includegraphics[width=\textwidth]{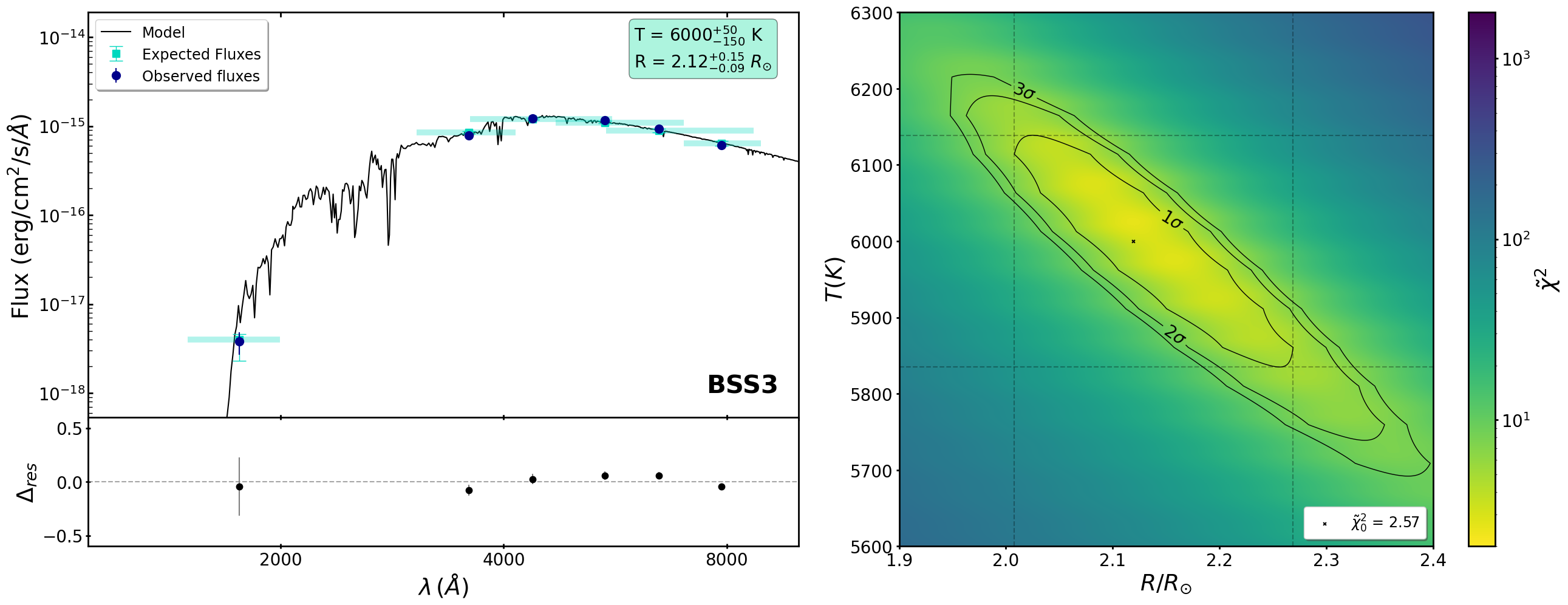}
     \caption{Same as in Figure~\ref{bss1_fit}, but for BSS3.}
     \label{bss3_fit}
 \end{figure*}
As expected from its CMD position, the best-fit temperature is rather low, and the radius is significantly larger than the values obtained for the other BSSs in the sample (see left panel of Fig. \ref{bss3_fit}). 
Both these values are reasonably in agreement with those determined from photometry (Table \ref{photo_estimates}), and also in this case there is no evidence of UV excess in the observed SED.
Since this BSS is the most CO-depleted depleted in our sample, and it shows no variability or W-UMa-type behaviour, we may suppose that its companion is an already cooled off WD.

\subsection{BSS4}
\begin{figure*}
    \centering
    \includegraphics[width=\textwidth]{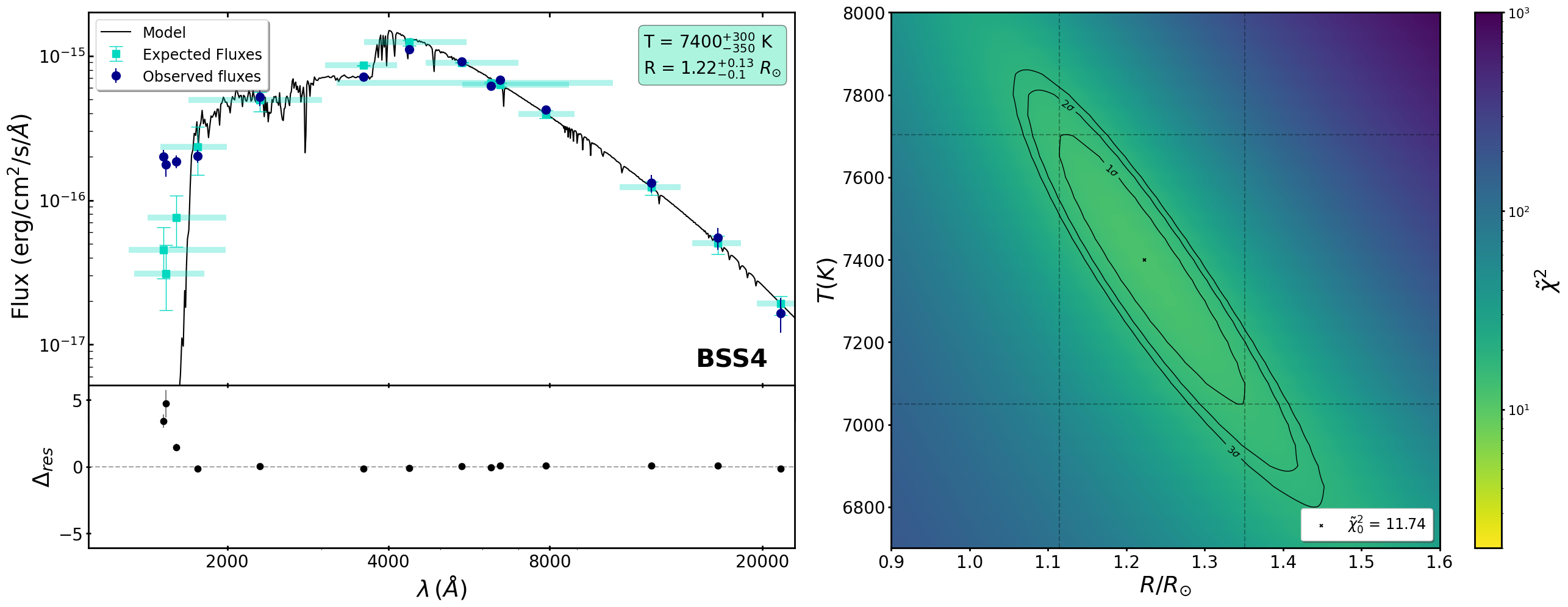}
     \caption{Same as in Figure~\ref{bss1_fit}, but for BSS4.}
     \label{bss4_fit}
 \end{figure*}

For BSS4, the photometry ranges from far-UV up to NIR wavelengths, allowing for the most complete SED reconstruction of the sample, with not only SBC, but also GALEX measurements available in the UV domain. In this case, the model SED for an isolated BSS is unable to reproduce the observed data points (Fig. \ref{bss4_fit}, left panel), clearly underestimating the observed far-UV measurements even for a best-fit effective temperature ($T=7400$ K) much larger than the photometric result ($T=6550$ K). Indeed, the minimum $\chi^2$ is very large (11.51) and the 1$\sigma$ region is wider than for the other BSSs, both in radius and in temperature, as the model tries to find suitable solutions for the UV flux. The fitting radius appears significantly smaller than the photometric estimate,
and it is a direct consequence of the model relaxing towards a hotter and more compact object in order to explain the far-UV photometry, whereas the photometric estimates rely on optical data.
Although this star has never been classified as a binary in previous works focussed on 47 Tucanae \citep[see, e.g.][]{Albrow2001,Weldrake2004}, the observed UV excess could be the photometric signature of the presence of a hot WD companion (see Sect. \ref{sec:wd_fitting}).

\subsection{BSS5}
\begin{figure*}
    \centering
    \includegraphics[width=\textwidth]{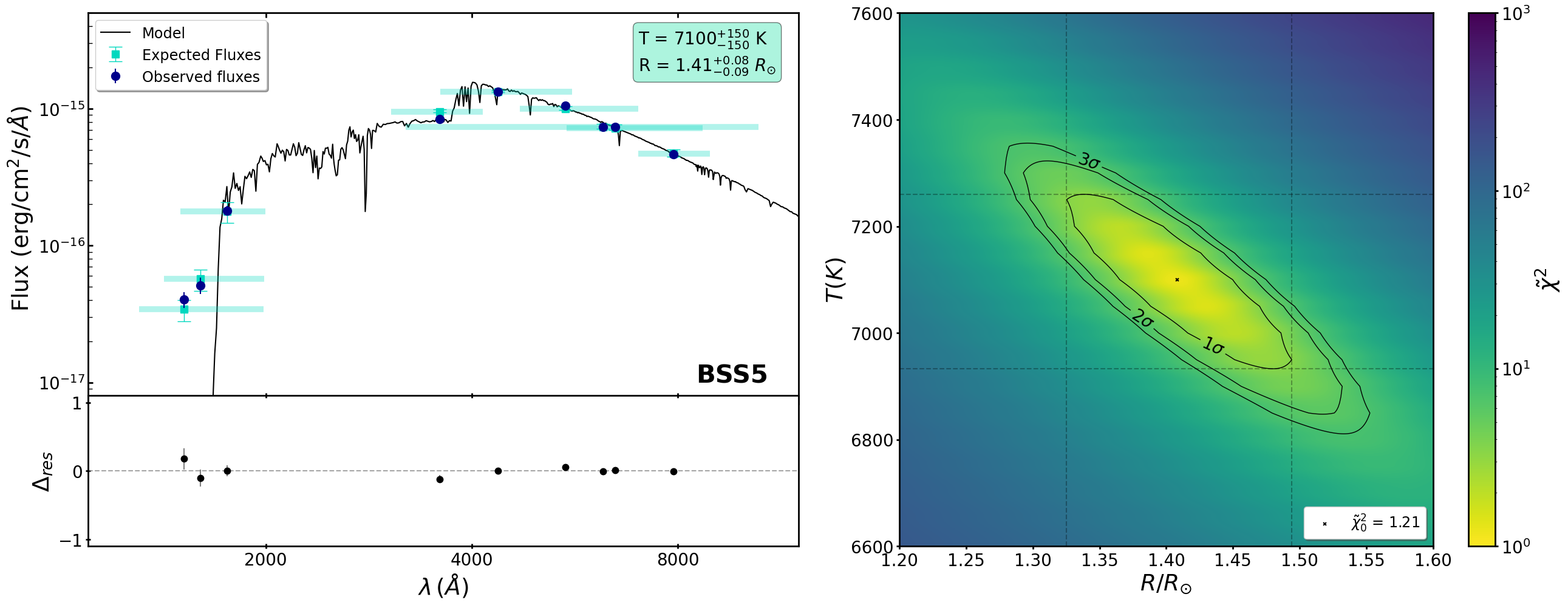}
     \caption{Same as in Figure~\ref{bss1_fit}, but for BSS5.}
     \label{bss5_fit}
 \end{figure*}

As shown in Fig. \ref{bss5_fit}, the whole observed SED of BSS5 can be properly reproduced by a model with stellar parameters ($T=7100$ K, $R=1.41 R_\odot$)
that are in reasonable agreement with the photometric estimates (Table \ref{photo_estimates}) and well constrained by the fit (see the 1$\sigma$ contour in the $\chi^2$ map). 
We conclude that this BSS is not orbited by a hot companion.

\begin{figure*}
    \centering
    \includegraphics[width=\textwidth]{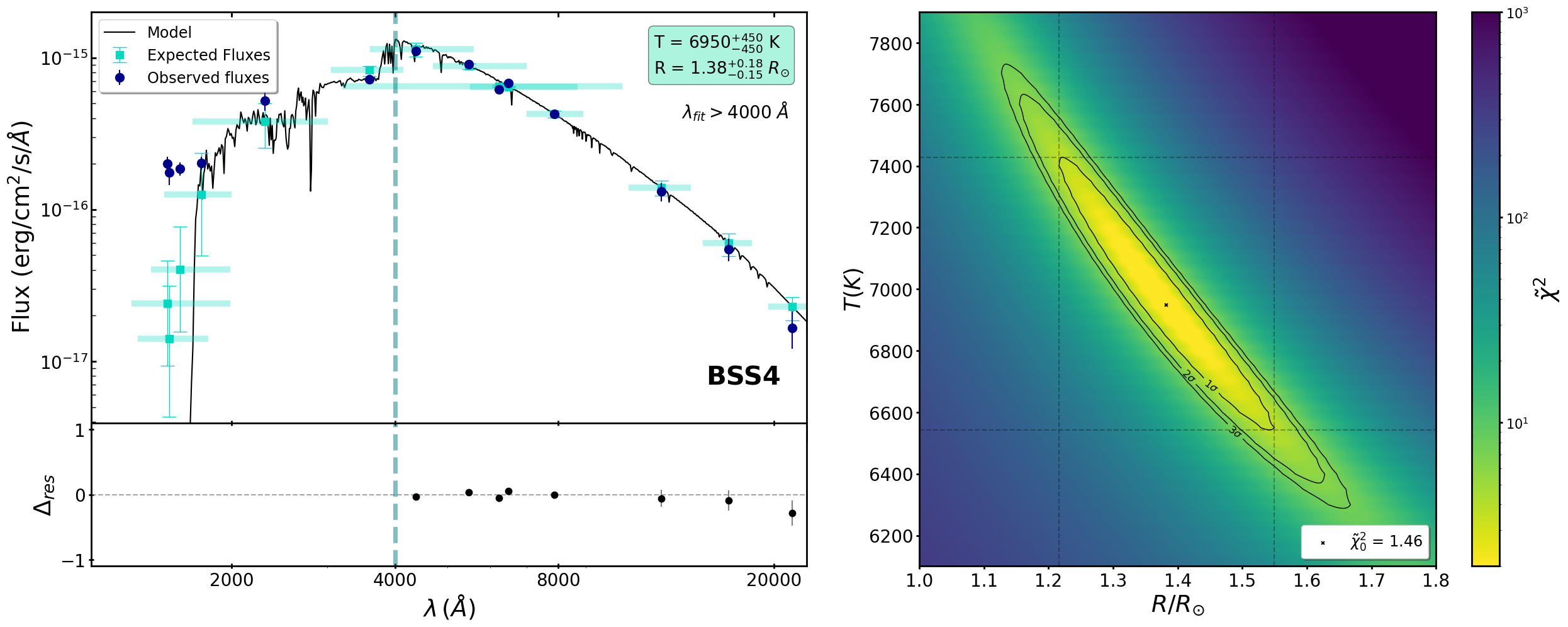}
     \caption{Same as in Figure~\ref{bss1_fit}, but for BSS4 and for the SED fitting procedure applied only to the photometric points with $\lambda > 4000$ \AA.}
     \label{bss4_fit_lcond}
 \end{figure*}

\begin{figure*}[htbp]
    \centering
    \begin{subfigure}[b]{0.49\textwidth}
        \centering
        \includegraphics[width=\textwidth]{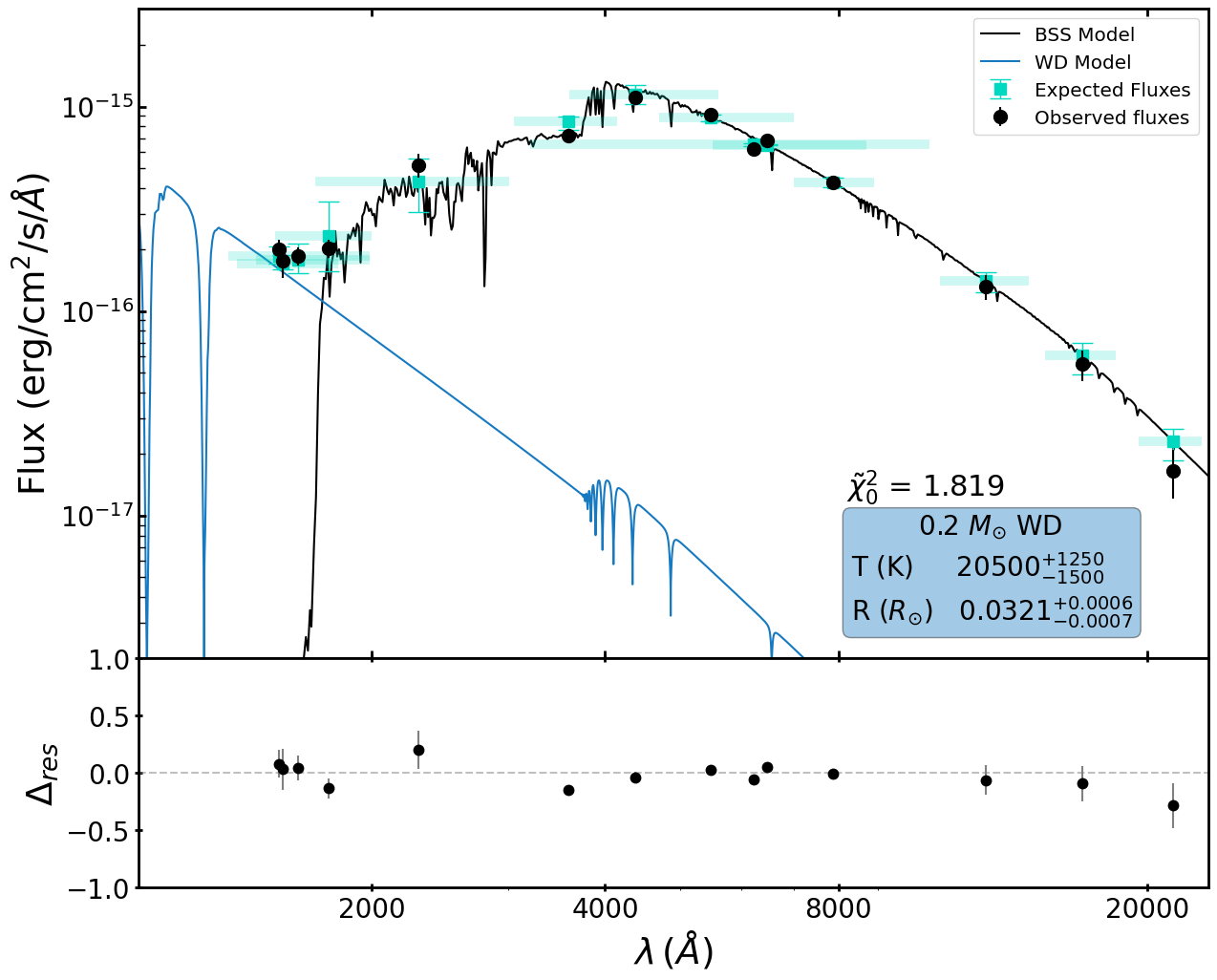}
    \end{subfigure}%
    \hfill
    \begin{subfigure}[b]{0.49\textwidth}
        \centering
        \includegraphics[width=\textwidth]{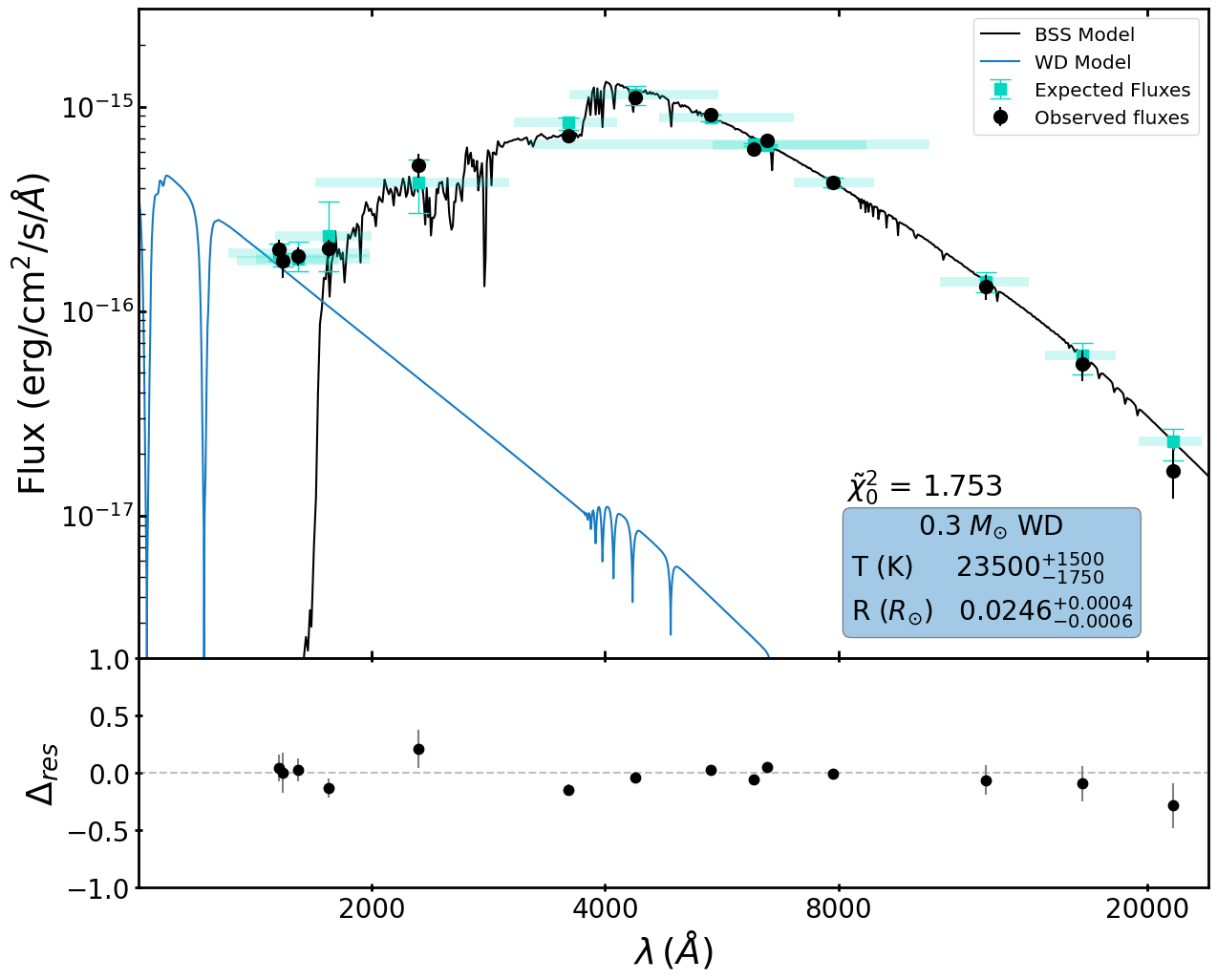}
    \end{subfigure}
    
    \vskip\baselineskip
    
    \begin{subfigure}[b]{0.49\textwidth}
        \centering
        \includegraphics[width=\textwidth]{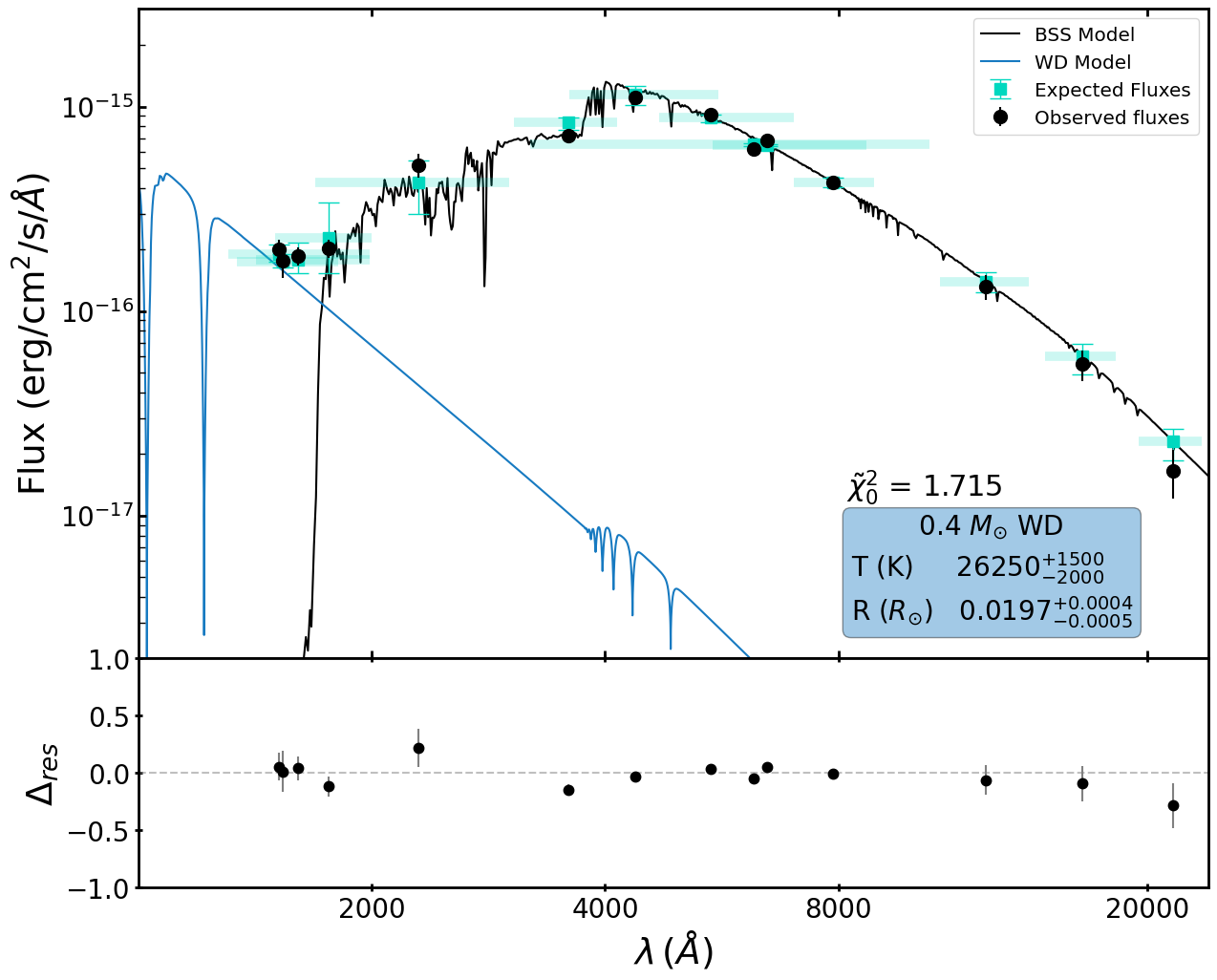}
    \end{subfigure}%
    \hfill
    \begin{subfigure}[b]{0.49\textwidth}
        \centering
        \includegraphics[width=\textwidth]{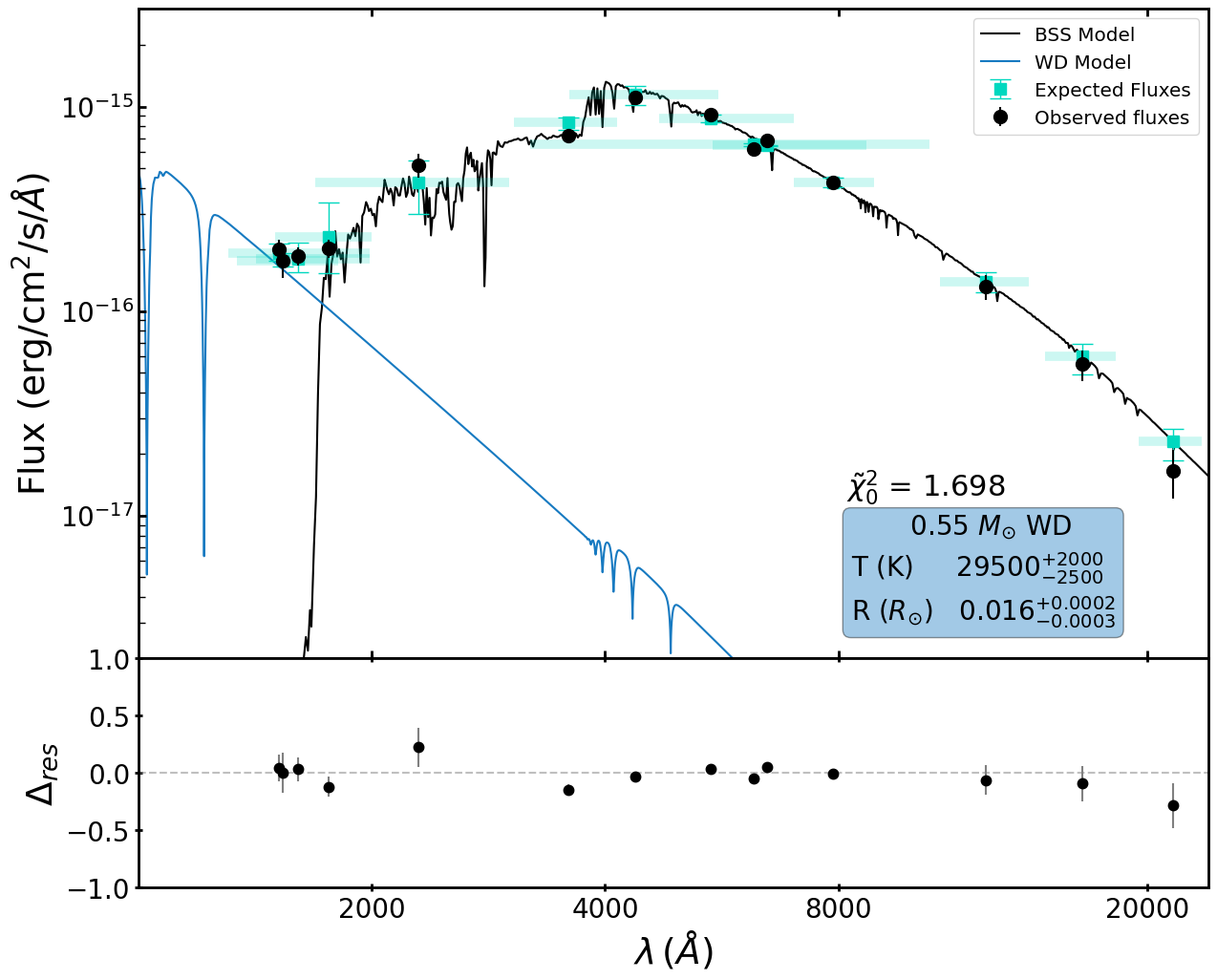}
    \end{subfigure}
    \caption{
    SED fitting of the BSS4 observed photometry (black circles, the same as in Figs. \ref{bss4_fit} and \ref{bss4_fit_lcond}) through the combination of the SED that best-fits the points at $\lambda > 4000$ \AA\ (black line, the same as in Fig. \ref{bss4_fit_lcond}) and the SED of different WDs (blue lines): from top-left to bottom-right, a He-WD with a mass of 0.2, 0.3, and $0.4 M_\odot$, and a $0.55 M_\odot$ CO-WD. The expected fluxes, computed from the convolution of the combined (BSS+WD) best-fit synthetic SED with the adopted photometric filters, are shown as cyan squares, and their error bars and horizontal extension have the same meaning as in the previous figures. The best-fit values of the WD surface temperature and radius are labelled, together with their $1\sigma$ uncertainty.}
    \label{ELM_models}
\end{figure*}

\section{WD companion to BSS 4} 
\label{sec:wd_fitting}
The unmistakable UV excess detected in BSS4 (Fig.\ref{bss4_fit}) cannot be explained in any way by the SED of an isolated BSS and points clearly to the presence of a hot companion, likely a hot WD.  
To constrain the properties of the latter, we searched for the combination of a BSS and a WD SED that best reproduces the observed photometric data. Albeit unresolved, the components of this binary system are expected to have the peak of their black-body emission in different regions of the spectrum. In particular, we expect the WD to be dominant in the far-UV domain, to contribute significantly at NUV wavelengths, and to be completely negligible above $\lambda\sim 4000$ \AA, where the BSS is the only contributor. 
Hence, we first searched for the model most appropriately describing the emission of the primary star alone (the BSS) by applying the SED fitting procedure described above only to the photometry at $\lambda \geq 4000$ \AA.
The resulting best-fit model converges towards $T=6950$ K and $R = 1.38~R_{\odot}$, 
(see left panel of Fig. \ref{bss4_fit_lcond}). 

To determine the properties of the companion star we would need to know the WD mass, because the temperature evolution and the radius of these compact, degenerate objects strictly depend on their mass. However, BSS4 has never been classified as a binary, and we therefore have no information about the WD mass. For this reason, we explored the cooling sequences of different mass/radius relations that are appropriate for old stellar systems. 
We started by considering which kind of donor star could leave the observed CO depletion on the BSS surface. In a GC such as 47 Tucanae, where the MS-TO mass is $\sim 0.85 M_\odot$ \citep{ferraro_16}, all stars along the MS and any post-MS evolutionary stages are experiencing or experienced core hydrogen burning through the proton-proton chain. As such, the CO-depleted material now observed on the BSS surface must have been generated in a shell surrounding the inactive core, where hydrogen thermonuclear burning occurs through the CNO cycle. Hence, the donor star must have been at least in the sub-giant branch (SGB). During this evolutionary stage, low-mass stars develop a He-core of at least $\sim 0.2 M_\odot$, which then increases at most up $\sim 0.5 M_\odot$ at the tip of the red giant branch (RGB). Hence, $0.2 M_\odot$ can be considered as a lower limit to the mass of the BSS companion that, being the stripped core of a SBG/RGB star, would be a He-WD (see also \citealt{cadelano19,cadelano20c,chen23}). If we consider, instead, the possibility of a CO-WD companion originated by MT activity from an asymptotic giant branch (AGB) star, its mass is expected to be around $\sim 0.55 M_\odot$ \citep{GellerMathieu2011}.
We therefore considered these two possibilities for the WD companion to BSS4, and we worked with temperature-radius (T-R) relations computed for He-WDs with masses $M = 0.2, 0.3, 0.4 M_\odot$ \citep[from][]{Bedard2020}, and a custom $0.55 M_\odot$ relation (calculated with the same code and physics inputs of \citealt{bastiwd} models) for the CO-WD scenario. 
To determine the WD fluxes, we used 1D pure-hydrogen (DA) LTE WDs spectral models computed by \citet{claret+20}, with temperatures ranging in the interval $4000$ K $< T < 40,000$ K in steps of $250$ K and surface gravities covering the $5.5 < \log{g} < 9.5$ interval in steps of $0.25$ dex. These models are typically used to describe the remnants found in short-period binary systems \citep{tremblay+15} and, as a consequence, they are ideal to treat WDs that did not follow canonical evolutionary paths, as in this case. 
We then performed a fitting procedure following the same steps described in Sect. \ref{sec:sed}, except that we constrained the T-R pairs to vary according to the adopted WD relations for fixed mass. The value of $\log{g}$ was chosen as the one closest to that provided by the T-R relations in the interval covered by \citet{claret+20} models. 
To find the best-fit WD model, the photometric data have been compared with the theoretical fluxes obtained from the sum of the WD and the BSS SEDs, after convolution with each photometric filter. The results are shown in Fig.~\ref{ELM_models}, while the derived WD properties are reported in Table~\ref{wd_bestfit_params} for the adopted mass values.
The fits return similar $\tilde{\chi}_0^2$ for all scenarios, showing that an object hotter than approximately 20,000 K is needed to properly reproduce the SED observed for this object.

\begin{table}[h!]
\caption{Best-fit parameters for WDs of different masses.}
\label{tab2}
\centering
\setlength{\tabcolsep}{5.pt}
\renewcommand{\arraystretch}{1.7}
    \normalsize
    \begin{tabular}{c c c c c c}
    \hline\hline
    $ M/M_\odot$ &  T/K  & $10^{-2} R/R_\odot$ & $\log g$ & $\tilde{\chi}_0^2$  
    & Age (Myr)\\
    \hline
    \multicolumn{6}{c}{He-WD} \\
    \hline
    0.2  &  $20500_{-1500}^{+1250}$ & $3.21_{-0.07}^{+0.06}$ & 6.75 & 1.819 & $0.11_{-0.05}^{+0.07}$ \\
    0.3  &  $23250_{-1750}^{+1500}$ & $2.46_{-0.06}^{+0.04}$ & 7.25 & 1.753 & $0.70_{-0.13}^{+0.99}$ \\
    0.4  &  $26250_{-2000}^{+1500}$ & $1.97_{-0.05}^{+0.04}$ & 7.5 & 1.715 & $8.11_{-2.14}^{+3.71}$ \\
    \hline
    \multicolumn{6}{c}{CO-WD} \\
    \hline
    0.55 &  $29500_{-2500}^{+2000}$ & $1.6_{-0.03}^{+0.02}$ & 7.75 & 1.698 & $12.19_{-2.10}^{+3.62}$ \\ 
    \hline
    \end{tabular}
    \label{wd_bestfit_params}
\end{table}

\section{Discussion and Conclusions}
\label{sec:conclusions}
This paper presents the search for the photometric signature of the MT process in a sample of 5 BSSs in 47 Tucanae, for which the spectroscopic signature had been previously detected (\citetalias{Ferraro06}). The analysis of far-UV photometric data acquired with the ACS/SBC at HST unambiguously reveals the presence of a hot companion to BSS4. This is the first clear-cut observational evidence of the link between CO-depletion and the presence of a hot companion, solidly confirming that these are both signatures of the MT process.

Both these characteristics are likely transient features, because rotational and internal mixing are expected to reduce or even erase any surface chemical anomaly. Although the time-scales of these processes are still unknown (and in some cases they can be of even a few Gyr; e.g., \citealp{thompson+08, stancliffe_glebbeek08}), the chemical investigation of large samples of BSSs in Galactic GCs suggests that they are much shorter than BSS lifetime. For instance, in the sample analysed by \citetalias{Ferraro06}, the CO signature was detected in only 5 BSSs out of 43 objects, corresponding to approximately  12\% of the total, suggesting that the chemical anomalies are rapidly erased during the BSS evolution. On the other hand, once the core of the stripped companion star becomes a WD, it is destined to progressively fade and cool. Hence, the detection of a modest or null UV flux (consistent with the SED of a BSS without companions) cannot be used as a solid argument in favour of an isolated BSS. Unfortunately, the large distance of 47 Tucanae \citep[$4.55\pm0.01$ kpc,][]{Ferraro1999} strongly limits our capability of pinpointing cool WDs in the UV. To evaluate the sensitivity of the acquired observations to the UV emission from faint (cool) WDs at the distance of 47 Tucanae, we performed a series of simulations convolving the SED of the observed BSSs with the SED of WDs with decreasing surface temperature (corresponding to increasingly fainter luminosity). The result varies according to the BSS temperature: for the coolest BSS in our sample (BSS2), our observations are not able to detect any UV flux emitted by WDs cooler than 12,000 K, while, for the hottest BSS in our sample (BSS5), the limit for a hidden WD is around 18,000 K. Hence, the non-detection of UV excess in some of the investigated (CO-depleted) BSSs may indicate the presence of a WD cooler than these thresholds. 
On the other hand, according to the working hypothesis presented by \citetalias{Ferraro06}, the  CO-depletion region in the [O/Fe]$-$[C/Fe] diagram (see Fig. \ref{fig:COdep}) is populated by systems at different stages of the MT process,  including stars that have just started, completed, or are still undergoing MT. Thus the non detection of a hot companion in the case of the other four BSSs in the sample, may indicate the presence of an old (cold) WD, as in the case of BSS2 and BSS3, or the fact that the MT process has not completed yet and it has not yet fully exposed the core of the donor star. This could be the case of BSS1 and BSS5, which are both classified as W Uma variables: BSS1 is the variable PC1-V10 of \citet{Albrow2001}, and BSS5 is V6 discussed in \citet{Weldrake2004} or E15 in \citet{kaluzny+13}. Indeed, the light curve of BSS5 shows secondary variations indicating that MT is still active.

Under the hypothesis that the strong UV excess observed for BSS4 is due to a hot WD, some consideration can be advanced about the properties of this binary system and its evolution. As discussed in Sect.~\ref{sec:wd_fitting}, the properties of the WD could be slightly different depending on the evolutionary stage of the donor star at the beginning of the MT process. If MT started at any stage between the SGB and the RGB tip, a degenerate He-core with a mass between 0.2 and 0.5 $M_\odot$ is expected. Conversely, if the donor star was evolving along the AGB, we expect a degenerate CO-core of approximately $0.55 M_\odot$. Of course, the total number of SGB and RGB stars in the cluster is much larger than that of AGB stars, suggesting that the first option should be more probable. However, since the activation of the MT process strictly depends on the parameters of the the binary system, it is not possible to exclude any alternative a priori. In the following we will therefore keep considering both options.

\begin{figure}
    \centering
    \includegraphics[width=0.5\textwidth]{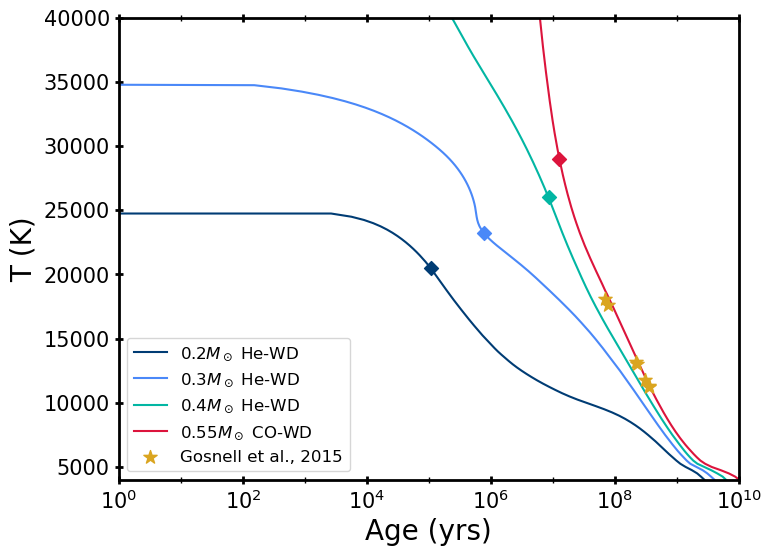}
     \caption{Cooling sequences of the different WD models (see the legend). The diamonds indicate the best fit-solutions obtained for the WD companion to BSS4 for each model. The golden stars indicate the CO-WD companions to BSSs found by \citet{Gosnell_2015} in NGC 188, for which the spectroscopically estimated mass is around $\sim 0.5 M_{\odot}$.}
     \label{T-age}
 \end{figure}

WDs of different masses evolve at different rates along different cooling sequences. Hence, the observed UV flux corresponds to different WD temperatures and cooling times depending on the WD mass. For the mass range appropriate for the investigated case (0.2 and 0.55 $M_\odot$) a quite hot WD, with temperature between $\sim 20,000$ K and $30,000$ K, is needed to account for the UV emission observed at the location of BSS4 (see Table \ref{wd_bestfit_params}). The corresponding cooling times range between less than 1 Myr (approximately 0.1 Myr) and $\sim 12$ Myr only (see Fig.~\ref{T-age} and Table \ref{wd_bestfit_params}). Hence, the relatively high temperature needed to reproduce the SED observed in BSS4 suggests that this is the result of a very recent MT event, which may not be entirely completed yet.

In Fig.~\ref{T-age} we show the cooling sequences for each of the investigated WD scenarios and the corresponding best-fit position of the BSS4 companion. For comparison, we also mark the position of the WD companions to BSSs found by \citetalias{Gosnell_2015} in the open cluster NGC 188 using SBC filters, for which the authors estimated masses around $\sim0.5 M_{\odot}$ from the binary parameters obtained from the observed spectra. Even when considering BSS+WD system detections performed using UVIT/AstroSat \citep{sahu+19,pal+24} or {\em Swift}/UVOT \citep{sheikh+24a,sheikh+24b}, BSS4 is one of the youngest BSS+WD systems observed so far, regardless of the adopted WD-mass scenario.

A high rotational velocity is also considered as a signature of recent formation, because the BSS is spun up by angular momentum transfer during the MT \citep[e.g.][]{packet81} and later slowed down by still unclear breaking processes such as disk locking or magnetic breaking \citep[see, e.g.,][and references therein]{sun+24}. Interestingly, the rotational velocity of BSS4 ($v \sin i=21$ km s$^{-1}$) is significantly larger than the median value (below 10 km s$^{-1}$) measured for the entire sample of 43 BSSs investigated by \citetalias{Ferraro06} (see also Fig. 1b in \citealt{ferraro23b}). Although this velocity is lower than the reference threshold adopted by \citet{ferraro23b} to select fast rotating (hence, likely young) BSSs, it is the largest in the sub-sample investigated here, and it could be explained by a low inclination angle of the rotation axis of BSS4 with respect to the line of sight, or a transition phase of the MT process. Thus, both the very young cooling age of the WD companion to BSS4 and its significant rotational velocity are consistent with the hypothesis that this is a recently formed system, possibly on the verge of completing the MT process. If we caught a MT-BSS in an intermediate stage of its formation pattern, we may expect that in the next few million years the stripped companion will increase its surface temperature because of the complete loss of envelope around its nucleus. Also, the BSS rotation velocity will further increase, and a larger CO-depletion will appear on its surface. If the MT has already ended, the WD companion is destined to progressively fade, rotation should slow down, and surface chemical anomalies will be possibly erased by mixing processes. 

Although the suggested scenario represents just an over-simplified scheme, it certainly is an intriguing working hypothesis worth further exploration. In particular, the determination of the radial velocity curve of the system (and also of the BSSs with no evidence of UV excess) could add further precious details for the physical characterisation of these puzzling objects and their evolutionary processes, providing crucial constraints to the theoretical modelling of the MT formation channel.   

\begin{acknowledgements}
This work is part of the project Cosmic-Lab at the Physics and Astronomy Department “A. Righi” of the Bologna University (\url{http:// www.cosmic-lab.eu/ Cosmic-Lab/Home.html}).
ER acknowledges the support from the European Union (ERC-2022-AdG, \say{StarDance: the non-canonical evolution of stars in clusters}, Grant Agreement 101093572, PI: E. Pancino). Views and opinions expressed are however those of the author(s) only and do not necessarily reflect those of the European Union or the European Research Council. Neither the European Union nor the granting authority can be held responsible for them. A.M. acknowledges support from the project "LEGO – Reconstructing the building blocks of the Galaxy by chemical tagging" (P.I. A. Mucciarelli), granted by the Italian MUR through contract PRIN 2022LLP8TK\_001.
\\
\indent This work made use of the following Python packages: Astropy\footnote{\citet{astropy:2013, astropy:2018, astropy:2022}}, numpy\footnote{\citet{harris2020array}}, pandas\footnote{\citet{reback2020pandas,mckinney-proc-scipy-2010}}, scipy\footnote{\citet{virtanen2020scipy}}, matplotlib\footnote{\citet{Hunter:2007}}, uncertainties\footnote{\href{https://pythonhosted.org/uncertainties/}{Uncertainties: a Python package for calculations with uncertainties}}, colour\footnote{\href{https://www.colour-science.org/}{Colour: Colour Science for Python}}.
This research has made use of the Spanish Virtual Observatory  project\footnote{\href{https://svo.cab.inta-csic.es}{SVO}} funded by MCIN/AEI/10.13039/501100011033/ through grant PID2020-112949GB-I00, and of SAOImage DS9, developed by the Smithsonian Astrophysical Observatory\footnote{\href{{http://ds9.si.edu}}{SAOImage DS9: A tool for astronomical data visualisation}}.
\end{acknowledgements}

%
%
\bibliographystyle{aa} 
\bibliography{aa56218-25} 

\newpage
\appendix
\onecolumn
\section{Photometric datapoints and parameters}
Table \ref{mag_table} lists all the data points used in the SED fitting procedure for each of the investigated BSS. In Table \ref{photflams}, we report the values of the extinction law \citep{Cardelli1989} and $photflam$ parameters adopted to convert magnitudes into fluxes in each filter, as described in Section \ref{sec:sed} \citep[see SVO Filter Profile Service,][]{SVO_2012,SVO_2020,SVO_2024}. 

\begin{table*}[h!]
    \centering
    \begin{threeparttable}
    \caption[Magnitudes of the BSSs of the sample]{Magnitudes of the target BSSs collected from the several photometric studies of 47 Tucanae at different wavelengths.}
    \small
    \setlength{\tabcolsep}{15.pt}
    \renewcommand{\arraystretch}{1.3}
    \begin{tabular}{c c c c c c }
    \hline
    \hline
     & BSS1 & BSS2 & BSS3 & BSS4 & BSS5 \\
    \hline
    \multicolumn{6}{c}{UBVRI filters} \\
    \hline
       U\tnote{a} & ... & ... & $16.89\pm0.01$ & $16.99\pm0.01$ & $16.82\pm0.02$ \\
       B\tnote{a} & ... & ... & $16.96\pm0.01$ & $17.07\pm0.01$ & $16.87\pm0.01$ \\
       V\tnote{a} & ... & ... & $16.40\pm0.01$ & $16.66\pm0.01$ & $16.50\pm0.01$ \\
       R\tnote{a} & ... & ... & $16.04\pm0.01$ & $16.39\pm0.01$ & $16.31\pm0.02$ \\
       I\tnote{a} & ... & ... & $15.71\pm0.01$ & $16.11\pm0.01$ & $16.01\pm0.01$  \\
    \hline
    \multicolumn{6}{c}{HST filters} \\
    \hline
       F225W\tnote{b} & $18.09\pm0.01$ & $18.37\pm0.01$ & ... & ... & ... \\
       F275W\tnote{c} & $17.48\pm0.04$ & $17.71\pm0.01$ & ... & ... & ... \\
       F300X\tnote{d} & $17.54\pm0.01$ & ... & ... & ... & ... \\
       F336W\tnote{c} & $16.90\pm0.04$ & $16.99\pm0.01$ & ... & ... & ... \\
       F435W\tnote{c} & $17.15\pm0.04$ & $17.24\pm0.01$ & ... & ... & ... \\
       F606W\tnote{c} & $16.68\pm0.01$ & $16.65\pm0.01$ & ... & ... & ... \\
       F814W\tnote{c} & $16.14\pm0.03$ & $16.20\pm0.01$ & ... & ... & ... \\
       F110W\tnote{e,f} & $15.94\pm0.01$ & $15.94\pm0.01$ & ... & ... & ... \\
       F125W\tnote{e} & $15.99\pm0.01$ & $15.84\pm0.01$ & ... & ... & ... \\
       F160W\tnote{f} & $15.69\pm0.03$ & $15.64\pm0.02$ & ... & ... & ... \\
      \hline
      \multicolumn{6}{c}{\text{GALEX, GAIA and 2MASS filters}} \\
      \hline
      fuv\tnote{g} & ... & ... & ... & $21.37\pm0.11$ & ... \\
      nuv\tnote{g} & ... & ... & ... & $19.35\pm0.06$ & ... \\
     G\tnote{h} & ... & ... & ... & $16.62\pm0.01$ & $16.44\pm0.01$ \\   
     J\tnote{i} & ... & ... & ... & $15.98\pm0.14$ & ... \\
     H\tnote{i} & ... & ... & ... & $15.81\pm0.17$ & ... \\
     K\tnote{i} & ... & ... & ... & $16.05\pm0.29$ & ... \\
     \hline
    \end{tabular}
    \begin{tablenotes}
        \item [a] \citet{Stetson2019}, VegaMAG
        \item [b] \citet{Rivera_Sandoval_2020}, VegaMAG
        \item [c] \citet{Nardiello2018}, VegaMAG
        \item [d] \citet{Cadelano2015}, VegaMAG
        \item [e] \citet{Pantoja_2018}, VegaMAG
        \item [f] \citet{Brown_2009}, VegaMAG
        \item [g] \citet{Dalessandro2012}, ABMAG
        \item [h] \citet{gaia_23}, VegaMAG
        \item [i] \citet{Cutri2003}, VegaMAG
    \end{tablenotes}
    \label{mag_table}
    \end{threeparttable}
\end{table*}

\begin{table}[h!]
\caption{Parameters adopted for the magnitude-to-flux conversion for each photometric filter.}
\label{tab2}
\centering
\setlength{\tabcolsep}{10.pt}
\renewcommand{\arraystretch}{1.6}
    \normalsize
    \begin{tabular}{c c c }
    \hline
    \hline
    \multirow{2}{*}{Filter} & \multirow{2}{*}{$A_{\lambda}/A_V$} & PHOTFLAM\\
    & & ($10^{-9}$ erg $cm^2$/s/\AA/$e^-$) \\
    \hline
    \multicolumn{3}{c}{HST filters} \\
    \hline
    F140LP & 2.64 & 6.5242 \\
    F150LP & 2.56 & 6.83489 \\
    F165LP & 2.52 & 6.43768 \\
    F225W & 2.64 & 4.2830 \\
    F275W & 2.03 & 3.78914 \\
    F300X & 1.93 & 3.75964 \\
    F336W & 1.69 & 3.30879 \\
    F435W & 1.32 & 6.45457 \\
    F606W & 0.924 & 2.86475 \\
    F814W & 0.602 & 1.1304 \\
    F110W & 0.34 & 0.400095 \\
    F125W & 0.3 & 0.299099 \\
    F160W & 0.22 & 0.142481 \\
    \hline
    \multicolumn{3}{c}{Stetson UBVRI} \\
    \hline
    U & 1.62 & 3.7282 \\
    B & 1.30 & 6.4227 \\
    V & 1.00 & 3.74234 \\
    R & 0.83 & 2.22727 \\
    I & 0.62 & 1.0992 \\
    \hline
    \multicolumn{3}{c}{GALEX, GAIA and 2MASS filters} \\
    \hline
    fuv & 2.63 & 46.0527 \\
    nuv & 2.86 & 20.5634 \\
    G & 0.87 & 2.49769 \\
    J & 0.305 & 0.3129 \\
    H & 0.193 & 0.1133 \\
    K & 0.125 & 0.0428 \\
    \hline
    \end{tabular}
    \label{photflams}
\end{table}

%
\end{document}